# Meaning vs. Information, Prediction vs. Memory, and Question vs. Answer

Yoonsuck Choe[1,2]
[1] Samsung Research
[2] Department of Computer Science and Engineering, Texas A&M University

## Abstract

Brain science and artificial intelligence have made great progress toward the understanding and engineering of the human mind. The progress has accelerated significantly since the turn of the century thanks to new methods for probing the brain (both structure and function), and rapid development in deep learning research. However, despite these new developments, there are still many open questions, such as how to understand the brain at the system level, and various robustness issues and limitations of deep learning. In this informal essay, I will talk about some of the concepts that are central to brain science and artificial intelligence, such as information and memory, and discuss how a different view on these concepts can help us move forward, beyond current limits of our understanding in these fields.

## 1 Introduction

Brain and neuroscience, psychology, artificial intelligence, all strive to understand and replicate the functioning of the human mind. Advanced methods for imaging, monitoring, and altering the activity of the brain at the whole-brain scale are now available, allowing us to probe the brain in unprecedented detail. These methods include high-resolution 3D imaging (using both physical sectioning and optical sectioning), monitoring ongoing neural activity (calcium imaging), and altering the activation of genetically specific neurons (optogenetics). On the other hand, in artificial intelligence, deep learning based on decades-old neural networks research made exponential progress, and it is now routinely beating human performance in many areas including object recognition and game playing.

However, despite such progress in both fields, there are still many open questions. In brain science, one of the main question is how to put together the many detailed experimental results into a system-level understanding of brain function. Also, there is the ultimate question to understand the phenomenon of consciousness. In artificial intelligence research, especially in deep learning, there are lingering issues of robustness (for example, deep neural networks were found to be easily fooled by slightly altered inputs such as adversarial inputs), and interpretability (deep neural networks are basically a black-box, and humans do not understand why or how they work so well).

In this essay, I will talk about some of the concepts that are central to brain science and artificial intelligence, such as information and memory, and discuss how a slightly different view on these concepts can help us move forward, beyond current limits of our understanding in these fields.

The rest of this chapter is organized as follows. In section 2, I will discuss meaning vs. information, with an emphasis on the need to consider the sensorimotor nature of brain function. In section 3, I will talk about prediction and memory, in the context of synaptic plasticity and brain dynamics. In section 4, I will move on to a broader topic of question vs. answer, and discuss how this dichotomy is relevant to both brain science and artificial intelligence. Section 5 will include some further discussions, followed by conclusions.

## 2. Meaning vs. Information

The concepts of computing and information have fundamentally altered the way we think about everything, including brain function and artificial intelligence. In analyzing the brain and in building and interpreting intelligent artifacts, computing has become a powerful metaphor, and information has become the fundamental unit of processing and measurement. We think about the brain and artificial neural networks in terms of computing and information processing, and measure their information content. In this section, I will talk mostly about information.

First of all, what is information? We tend to use this word in a very loose sense, and information defined in this way is imbued with meaning (or semantic content). In a scientific/engineering discussion, information usually refers to the definition given in Claude Shannon's information theory[1]. In information theory, information is based on the probability of occurrence of each piece of message, and the concept is used to derive optimal bounds on the transmission of data.

However, there can be an issue if we try to use Shannon's definition of information in our everyday sense, since as Shannon explicitly stated in his primary work on information theory, information defined as such does not have any meaning attached to it. So, for example, in an engineered information system, when we store information or transmit information, the data themselves do not have meaning. The meaning only resides in the human who accesses and views the information. All the processing and transmission in the information system is at a symbolic level, not at a semantic level. Philosopher John Searle's paper on the "Chinese room argument"[2] made clear of the limitation of the computational/information processing view of cognition. Inside the Chinese room there is an English monolingual person with all the instructions for processing Chinese language. The room has two mail slots, one for input, and one for output. A Chinese speaker standing outside the room writes down something on a piece of paper and deposits it in the input slot, and the English speaker inside will process the information based on the instructions present in the room and draw (not write) the answer on a piece of paper and returns it through the output slot. From the outside, the Chinese room speaks and understands perfect Chinese, but there is no true understanding of Chinese in this system. The main problem is that the information within the room lack meaning, and this is why grounding is necessary; grounding in the sense that information is grounded in reality, not hovering above in an abstract realm of symbols (see Stevan Harnad's concept of symbol grounding[3]). Many current artificial intelligence systems including deep learning tend to lack such grounding, and this can lead to brittleness, since these systems simply learn the input output mapping, without understanding.

---

[1] "Category:Information Theory - Scholarpedia." 26 May. 2011, http://www.scholarpedia.org/article/Category:Information_Theory. Accessed 17 Dec. 2017.

Thus, we need to think in terms of the meaning of the information, how semantic grounding is to be done: How does the brain ground information within itself? How can artificial intelligence systems ground information within itself? What is the nature of such grounding? Perceptual? Referential? This can be a very complex problem, so let us consider a greatly simplified version. Suppose you are sitting inside a totally dark room, and you only observe the occasional blinking of some light bulbs. You count the bulbs, and it looks like there are four of them. Each of these bulbs seem to be representing some information, but you are unsure what they mean. So, here you have a classic symbol grounding problem. The light blubs are like symbols, and they represent something. However, sitting inside this room, it seems that there is no way you can figure out the meaning of these blinking lights. Now consider that the dark room is the primary visual cortex (V1), and the four light bulbs are neurons that represent something, and in place of you, we put inside the room the downstream visual areas. With our reasoning above, it would suggest that the downstream visual areas have no way to understand the meaning of V1 activities, which seems absurd: one should be blind.

It turns out that this problem can only be solved if we allow motor interaction from within the system. Inside the dark room, we can install a joystick, and the person sitting inside can move it around and see how the joystick movement relates to the changes in the blinking light in a systematic manner. Consider the case where the four light bulbs represent four different orientations 0°, 45°, 90°, and 135°, respectively. How can movement of the joystick reveal the meaning of these light bulbs? Suppose the joystick is linked to a camera external to the room and the direction of gaze of the camera follows the joystick control. In the external visual environment, assume there is a single long line that is oriented 45°, and that the camera is pointed toward one segment of the line. In this set up, the second light bulb (representing 45°) will be turned on. If the joystick is moved in a direction other than 45° and 225°, the lights will go off (note: if there were other lines in the environment, a different light will turn on). However, when the joystick is moved in these two directions (45° and 225°), the second light bulb will be kept turned on (i.e., it will remain invariant). In this case, the property of the second light bulb and the property of the movement that keeps the light invariant are exactly aligned. Through this kind of sensorimotor exploration, the property of the internal representation can be recovered, from within the system (without direct perceptual access to the external environment), thus the meaning can remain intrinsic to the system. In our lab, we explored these ideas in a reinforcement learning setting (learn a policy *p* that maps from state *S* [orientation] to action *A* [gaze direction]), where we showed that the internal state invariance criterion (the reward) can be used for motor grounding of internal sensory representation in a simple visuomotor agent. See [1] and subsequent works for more details.

---

2 "Chinese room argument - Scholarpedia." 26 Aug. 2009, http://www.scholarpedia.org/article/Chinese_room_argument. Accessed 17 Dec. 2017.
3 "Symbol grounding problem - Scholarpedia." 6 May. 2007, http://www.scholarpedia.org/article/Symbol_grounding_problem. Accessed 17 Dec. 2017.

To sum up, meaning is central to brain science and artificial intelligence, and to provide meaning to information, it is critical to consider the sensorimotor aspect of the information system, whether natural or artificial.

## 3. Prediction vs. Memory

Many questions in brain and neuroscience focus on the concept of plasticity, how the brain changes and adapts due to experience, and this leads to the question of memory. Connections between neurons adapt over time (synaptic plasticity: long term, short term, etc.), and ongoing neural dynamic of the brain can also be altered by the immediate input stimulus. On a higher level, plasticity is usually considered in relation to various forms of memory: long term memory, short term memory, working memory, episodic memory, implicit memory, explicit memory, etc. Also, in a common sense way, people ask how the brain remembers, and what constitutes memory in the brain. In artificial intelligence, the same is true: How information should be represented, stored, and retrieved; how connection weights should be adapted in artificial neural networks to store knowledge; and how neural networks can be used to utilize external memory, etc.

What is memory, and how is it related to prediction, and why should we think more about prediction than memory? Memory is backward looking, directed toward the past, while prediction is forward looking, and is directed toward the future. Memory enables prediction, since without memory, the system will be purely reactive, living in the eternal present. So, again, why should we direct our attention toward prediction? In terms of brain function and artifacts that try to mimic it, prediction is of prime importance. In our everyday life, moment to moment prediction and long term prediction play a critical role. Simple tasks as walking, navigating, and many daily activities involving interaction with the environment and with other humans require prediction. Long term predictions of phenomena such as seasonal changes enable planning and improved productivity. So, in a sense, prediction is an important brain function, and it is increasingly being recognized as a central function of the brain as well as a key ingredient in intelligent machines (for an overview of related ideas, see Andy Clark's paper [2], and various papers on the use of predicted future states in reinforcement learning).

In this section, I will talk about how such predictive function could have emerged in the brain, how it is related to synaptic plasticity mechanisms (memory), how it is relevant to the study of neural networks, and how predictive properties in the brain can be linked to higher level phenomena such as consciousness.

First, consider delay in the nervous system. Neurons send their signals to their receiving neurons via elongated wires called axons. Transmission through these axons can take few milliseconds (ms), the duration depending on various factors such as the length, diameter, and whether the axon is insulated with myelin or not. When you add up the delay, it comes to a pretty significant amount of time: about 180 ms to 260 ms from stimulus presentation to behavioral reaction [3]. This kind of delay may be considered bad for the system, since it can be a matter of life and death, especially for fast moving animals. Also, in engineering systems, delay is considered a great hindrance. However, delay can be useful in two ways: (1) in a reactive system such as a feedforward neural network, addition of delay in the input can effectively add memory, and (2) mechanisms evolved to counteract the adverse effects of delay can naturally lead to predictive capabilities.

In [4], we showed that addition of delay in feedforward neural network controller can solve a 2D pole balancing problem that does not include velocity input. Also, in a series of works we showed that certain forms of synaptic plasticity (dynamic synapses) can be considered as a delay compensation mechanism, and how it relates to curious perceptual phenomena such as the flash lag effect (see [5] for an overview). In flash lag effect, a moving object is perceived as being ahead of a statically flashed object that is spatially aligned. One explanation for this phenomenon is that the brain is compensating for the delay in its system, by generating an illusion that is aligned, in real time, with the current state of the external environment. For example, image of the two aligned objects hit the retina. The information takes several milliseconds to reach the visual area in the brain. In the meanwhile, one of the objects moves ahead, so by the time the two objects are perceived (when the information arrives in the visual area), in the environment, they are misaligned because the moving object has moved on. The argument is that flash-lag effect allows the brain to perceive this as misaligned objects, which is more in line with the actual environmental state at the time of perception. We showed that facilitating neural dynamics, based on dynamic synapses (the facilitating kind, not the depressing one: see Henry Markram and colleagues' works cited in [5]) can replicate this phenomenon, and furthermore, the use of spike-timing-dependent plasticity (STDP) can help explain more complex phenomena such as orientation flash-lag effect (see [5] and references within).

Second, we will consider predictive properties in brain dynamics and how it can be related to higher level phenomena such as consciousness. As we discussed above, prediction seems to be a key function of the brain. How can it also be used to gain insights into phenomena such as consciousness? In consciousness studies, the neural correlate is highly sought after, where neural correlates of consciousness refer to the "... neural events and structures … sufficient for conscious percept or conscious memory" [6]. This view is somewhat static (of course it depends on the definition of "event"), and its dependence on sufficient condition can lead to issues relating to the hard problem of consciousness — how and why it "feels" like it. In our view, it would be better to first consider the necessary condition of consciousness, and this led us to the realization that the property of brain dynamics, not just isolated "events", need to be considered. We also found that predictive property in brain dynamics has an important role to play in consciousness, and this is how the discussion of consciousness comes into picture in this section [7].

Let us consider necessary conditions of consciousness. We begin by considering consciousness and its subject. There cannot be a consciousness without a subject, since consciousness, being a subjective phenomena, cannot be subjective without a subject. Next, consider the property of the subject (or let us say self). Self is the author of its own actions, and there is a very peculiar property about these actions authored by the self — that it is 100% predictable. When I say "I will clap my hands in 5 seconds", in 5 seconds, I will make sure that happens, so that my behavior in such a case is 100% predictable. This is quite unlike most phenomena in the world that is not so much the case. In order to support such prediction, some part of the brain has to have a dynamic pattern that has a predictable property. That is, based on past data points in the neural dynamic trajectory, it needs to be possible to predict the current data point. This, we believe, is an important necessary condition of consciousness (see [7] for details). Through computational simulations and secondary analysis of public EEG data, we showed that predictive dynamics can emerge and have fitness advantage in synthetic evolution [7], and conscious states such as awake condition and REM sleep condition exhibit more predictive dynamics than unconscious states (slow-wave sleep) [8].

For the first study [7], we evolved simple recurrent neural network controllers to tackle the 2D pole-balancing task, and found that successful individuals have a varying degree of predictability in its internal dynamics (how the hidden unit activities change over time). This is discouraging, since if individuals with internal dynamics with high or low predictability are equally good in behavioral performance, predictive dynamics may not evolve to dominate. However, a slight change in the environment made individuals with high predictive dynamics to proliferate. The only change necessary was to make the task a little harder (make the initial random tilt of the pole to be more). This suggests that predictive internal dynamics have a fitness value when the environment changes over time, and this happens to be how the nature is, thus predictive dynamics will become dominant. Not the strongest or the fastest species survive: The most adaptable species survive, where prediction seems to be the key, and this lays down the necessary condition for consciousness.

In the second study [8], we analyzed publicly available brain EEG data collected during awake, rapid eye movement (REM) sleep, and slow-wave sleep. Since awake and vivid dreaming (REM sleep) are associated with consciousness while deep sleep (slow-wave sleep) with unconsciousness, we measured the predictability in these EEG signal wave forms. We preprocessed the raw EEG signal, computed the inter-peak interval (IPI), the time distance between peaks in the EEG signal, and measured how easy it is to predict the next IPI based on previous IPI data points. We found that awake and REM EEG signals have higher IPI predictability that that of slow-wave sleep, suggesting that IPI predictability and consciousness may be correlated.

In this section I discussed how synaptic plasticity mechanisms can be directly linked to prediction, how delay in the nervous system may have led to predictive capabilities, and how predictive dynamics can serve as a precursor of consciousness. In sum, prediction is a key function of the brain, and it should also be included as such in artificial systems.

## 4. Question vs. Answer

In both brain science and artificial intelligence, the general focus is to understand how the brain solves problems relating to perceptual, cognitive, and motor tasks, or how to make artificial intelligence algorithms solve problems in vision, natural language processing, game playing, robot control, etc. That is, we are focused on mechanisms that produce answers, and less on mechanisms that pose the questions. Of course we know the importance of asking the right questions, and any researcher is well aware of the importance of picking the right research question. Often times, research involves finding new ways to conceive of the problem, rather than finding new ways of solving problems as conceived [9], and this is especially essential when the conceived problem itself is ill-formed so as to be unsolvable (e.g., how can we prove Euclid's 5th postulate [unsolvable], vs. can we prove Euclid's 5th postulate [solvable]).

In 2012, Mann and I discussed in [10] the need to start paying attention to problem posing, as opposed to problem solving. It turns out that problem posing has been an active topic in the education literature (see [11] and many subsequent publications). So, learning and problem posing seems to be intricately related. However, this angle is not explored much in artificial intelligence, except for rare exceptions, and I strongly believe integrating learning and problem posing can lead to a much more robust and more general artificial intelligence. Some of those rare exceptions is Schmidhuber's study, which explicitly addresses this issue. In his Powerplay algorithm, both problems and solvers are parameterized and the algorithm seeks specific problems that are solvable with the current capability of the solver, and loop through this to train an increasingly general problem solver [12]. More recently, question asking has been employed in interactive problem solving in robotics [13] and vision problems [14]. However, these are done within a strict task framework, so open-ended questions or questions that question the validity of existing questions cannot be generated (see [15] for a bit more open-ended approach called inverse Visual Question Answering). For an intelligent agent, this latter form of questioning (or problem posing) will become increasingly important, as the current learning algorithms cannot easily go beyond its defined task context. Some ideas we discussed in [10] for problem posing include: (1) recognizing events in the environment that can be potentially become a problem to be solved, (2) checking if existing problems are ill-posed, and (3) given an overarching goal, come up with smaller problems that may be of different kind than the original goal (if they are of the same kind, straight-forward divide-and-conquer algorithms can be used).

How can the idea of question vs. answer be relevant to brain science? I think it is relevant since the topic has not received attention that it deserves, and question asking (or problem posing) is an important function of the brain. There are many papers on decision making, but not much on how the brain asks questions that requires subsequent decision making. Understanding the brain mechanism of question asking can lead to new discoveries regarding core brain function, and in turn the insight can help us build better intelligent artifacts.

In sum, question asking needs more attention from brain science and artificial intelligence, in order for us to gain a deeper understanding of brain function and to build more intelligent artifacts.

## 5. Discussion

In this chapter, I talked about several dichotomies of concepts that are important to brain science and artificial intelligence: meaning vs. information, prediction vs. memory, and question vs. answer, with an emphasis on the first concept in each pair. Below, I will discuss additional related works that have relevance to the topics I discussed in the preceding sections.

In terms of meaning, deep neural network research approaches the issue from a different angle than the one presented in this chapter — that of embedding, e.g., word and sentence embedding [16]. The main idea of embedding is to map words or sentences (or other raw input) into a vector space where concepts can be manipulated algebraically based on their meaning. For example, when vector representations of "Germany" and "capital" are added, the resulting vector represents "Berlin" (example from [16]). The main idea in this case is to train a neural network to map the input word to a series of words that appear before or after the input word in a sentence. The hidden layer representation then tends to have this desired semantic property. This is one step toward meaningful information. However, whether the meaning in this case is intrinsic to the system, i.e., interpretable from within the system, is an open question.

As we saw in section 2, motor aspect is important for semantic grounding. In a related context, philosopher Ludwig Wittgenstein proposed that the meaning of language is in its use [17]. Basically, this is a departure from meaning based on what it (e.g., a word) represents. A more recent thesis in this general direction comes from Glenberg and Robertson [18], where they emphasized that "what gives meaning to a situation is grounded in actions particularized for that situation", thus taking an action-oriented view of grounding. Also see O'Regan and Noë's sensorimotor contingency theory, which is organized around a similar argument [19].

One interesting question is, does the range of possible motor behavior somehow limit the degree of understanding? That is, can organisms with higher degree of freedom and richer repertoire of actions gain higher level of understanding? I believe this is true. For example, recall the orientation perception thought experiment in Section 2. If the visuomotor agent was only able to move horizontally or vertically, but not diagonally, it would never be able to figure out what the $45°$ and $135°$ light bulbs mean. Intelligence is generally associated with the brain size or brain/body ratio, but what may also be very important is how rich the behavioral repertoire of the animal is. For example, all the animals we consider to be intelligent have such flexibility in behavior: primates, elephants, dolphins, and even octopuses. An extension of this idea is, can an agent extend its behavioral repertoire? This is possible by learning new moves, but it is also possible by using tools. The degree of understanding can exponentially grow if the agent can also construct increasingly more complex tools. This I think is one of the keys to human's superior intelligence. See [20] for our latest work on tool construction and tool use in a simple neuroevolution agent, and our earlier work on tool use referenced within.

In section 2, I also proposed the internal state invariance criterion, within the context of reinforcement learning. This raises an interesting idea regarding rewards in reinforcement learning. In traditional reinforcement learning, the reward comes from the external environment. However, research in reinforcement learning started to explore the importance of rewards generated from within the learning agent. This is called "intrinsic motivation" [21], and the internal state invariance criterion is a good candidate. In this view, intrinsic motivation also seems to be an important ingredient for meaning that is intrinsic to the learning system. Another related work in this direction is [22], based on the criterion of independently controllable features. The main idea is to look for good internal representations where "good" is defined by whether action can independently control these representations or not. So, in this case, both the perceptual representations and the motor policy are learned. This kind of criterion can be internal to the agent, thus, keeping things intrinsic, while allowing the agent to understand the external environment. Also see [23] for our work on co-development of visual receptive fields (perceptual representations) and the motor policy.

Next, I would like to discuss various mechanisms that can serve as memory, and how, in the end, they all lead to prediction. In neural networks, there are several ways to make the network responsive to input from the past. Delayed input lines is one, which allows a reactive feedforward network to take input from the past into consideration when processing the current input (see e.g. [4]). Another approach is to include recurrent connections, connections that form a loop. More sophisticated methods exist such as Long Short Term Memory (LSTM), etc., but generally they all fall under the banner of recurrent neural networks. Finally, there is a third category that can serve as a memory mechanism, which is to allow feed forward neural networks to drop and detect token-like objects in the environment. We have shown that this strategy can be used in tasks that require memory, just using feed forward neural networks [24]. From an evolutionary point of view, reactive feedforward neural networks may have appeared first, and subsequently, delay and ability to utilize external materials may have evolved (note that this is different with systems that have an integrated external memory, e.g., Differentiable Neural Computers [25]). Further development or internalization of some of these methods (especially the external material interaction mechanism) may have led to a fully internalized memory. These memory mechanisms involve some kind of recurrent loop (perhaps except for the delayed input case), thus giving rise to dynamic internal state. As we have seen in section 3, in such a system, networks with predictive dynamics have a fitness advantage, and thus will proliferate.

Continuing with the discussion on prediction, I would like to talk about how recent advances in machine learning are benefitting from the use of prediction as a learning objective/criterion. In machine learning situations where explicit target values are rare or task-specific rewards are very sparse, it is a challenge to train effectively the learning model. Recent work by Finn and Levine [26] (and others) showed that learning motor tasks in a completely self-supervised manner is possible without a detailed reward, by using a deep predictive model, which uses a large data set of robotic pushing experiment. This shows a concrete example where prediction can be helpful to the agent. See [26] for more references on related approaches that utilize prediction.

Finally, let us consider question asking. As briefly hinted in Section 4 (citing [10]), generating questions or posing problems can be viewed as generating new goals. Similar in spirit with Schmidhuber's Powerplay [12], Held et al. proposed an algorithm for automatic goal generation in a reinforcement learning setting [27]. The algorithm is used to generate a range of tasks that the agent can currently perform. A generator network is used to propose a new task to the agent, where the task is drawn from a parameterized subset of the state space. A significant finding based on this approach is that the agent can efficiently and automatically learn a large range of different tasks without much prior knowledge. These results show the powerful role of question asking in learning agents.

## 6. Conclusion

In this chapter, I talked about meaning vs. information, prediction vs. memory, and question vs. answer. These ideas challenge our ingrained views of brain function and intelligence (information, memory, and problem solving), and we saw how the momentum is building up to support the alternative views. In summary, we should pay attention to (1) meaning, and how it can be recovered through action, (2) prediction as a central function of the brain and artificial intelligence agents, and (3) question asking (or problem posing) as an important requirement for robust artificial intelligence, and the need to understand question asking mechanisms in brain science.

**Acknowledgments**
I would like to thank Asim Roy and Robert Kozma, who, together with myself, chaired a panel discussion on "Cutting Edge Neural Networks Research" at the 30th anniversary International Joint Conference on Neural Networks in Anchorage, Alaska, where I had the opportunity to refine many of my previous ideas, especially on meaning vs. information. I would also like to thank the panelists Peter Erdi, Alex Graves, Henry Markram, Leonid Perlovsky, Jose Principe, and Hava Siegelmann, and those in the audience who participated in the discussion. The full transcript of the panel discussion is available at https://goo.gl/297j1d. Finally, I would like to thank my current and former students who helped develop and test the ideas in this chapter.